\documentclass[pop,fleqn]{w-art}
\usepackage{times}
\usepackage{w-thm}

\theoremstyle{plain}

\theoremstyle{definition}

\usepackage{graphicx}
\usepackage{amssymb}

\newcommand{\calT}{{\cal T}}

\newcommand{\calH}{{\cal H}}
\newcommand{\calW}{{\cal W}}

\providecommand{\pd}[2]{\frac{\partial #1}{\partial #2}}

\begin{document}

\DOIsuffix{theDOIsuffix}



\title[Transport dynamics of single ions in segmented microstructured Paul 
trap arrays]{Transport dynamics of single ions in segmented microstructured 
Paul trap arrays}

\author[R. Reichle et al.]{R. Reichle\inst{1,2}
  \footnote{Corresponding author\quad E-mail:
      ~\textsf{rainer.reichle@uni-ulm.de}}}

\author[ ]{D.~Leibfried\inst{2,}}
\author[ ]{R.~B.~Blakestad\inst{2,}}
\author[ ]{J.~Britton\inst{2,}}
\author[ ]{J.~D.~Jost\inst{2,}}
\author[ ]{E.~Knill\inst{2,}}
\author[ ]{C.~Langer\inst{2,}}
\author[ ]{R.~Ozeri\inst{2,}}
\author[ ]{S.~Seidelin\inst{2,}}
\author[ ]{D.~J.~Wineland\inst{2,}}

\address[\inst{1}]{University of Ulm, Albert-Einstein-Allee 11, D-89069 Ulm}
\address[\inst{2}]{National Institute for Standards and Technology, 
325 Broadway, Boulder/CO 80305}

\begin{abstract}
It was recently proposed to use small groups of trapped ions as qubit carriers 
in miniaturized electrode arrays that comprise a large number of individual 
trapping zones, between which ions could be moved \cite{bible,kielpinski}.  
This approach might be scalable for 
quantum information processing with a large numbers of qubits. Processing of 
quantum information is achieved by transporting ions to and from separate 
memory and qubit manipulation zones in between quantum logic operations. The 
transport of ion groups in this scheme plays a major role and requires precise 
experimental control and fast transport. In this paper we introduce a
theoretical framework to study ion transport in external potentials that might 
be created by typical miniaturized Paul trap electrode arrays. In particular 
we discuss the relationship between classical and quantum descriptions of the 
transport and study the energy transfer to the oscillatory motion during 
near-adiabatic transport. Based on our findings we suggest a numerical method 
to find electrode potentials as a function of time to optimize the local 
potential an ion experiences during transport.  We demonstrate this method
for one specific electrode geometry that should closely represent the 
situation encountered in realistic trap arrays. 
\end{abstract}
\maketitle                   

\section{Introduction}
Quantum information processing is a rapidly evolving field of physical 
science. Its practical importance arises from the exponential speedup in 
computation of certain algorithmic tasks over classical computation 
\cite{nielsen}. Building an actual device that can process quantum 
information, however, is technologically difficult due to the need for qubits 
that can be processed and read out with high fidelities and the extreme 
sensitivity of the quantum mechanical states stored in these units against 
external uncontrolled perturbations. A promising technical approach as shown 
over the last decade, is to use strings of ions as physical qubits confined in 
linear electromagnetic Paul traps \cite{bible,CiracZoller}. These strings are 
stored in a single trap and constitute a one dimensional crystallized 
structure whose vibrational modes can be laser cooled to their ground states. 
The strong mutual coupling of the ions by Coulomb forces  in such a crystal 
has been proposed and utilized to create arbitrary superpositions of quantum 
states of the ionic internal states (\cite{CiracZoller,Cat,AustrianCat}).
In the last few years methods were developed that enable quantum state 
engineering with high precision and long coherence times 
\cite{phasegate,memory, memory2, haljan, donnell}. The necessary criteria 
\cite{DiVincenzo} for 
large-scale quantum computation have been demonstrated in the past years, 
and small algorithms have been implemented successfully 
\cite{teleportation2,teleportation,QEC,QFT,Monroe}. However, as in other 
approaches aiming towards quantum computation, scaling to many qubits is 
challenging. Considerable overhead is required by quantum 
error correcting schemes that permit robust quantum computation and make 
large-scale implementations feasible. To scale up a linear string of many 
ions, a rapidly growing number of vibrational degrees of freedom needs to be 
controlled and cooled to the ground state for reliable processing. This is 
extremely difficult to realize. A more recent proposal \cite{bible,kielpinski} has 
been made to circumvent this problem by using small arrays of a few qubits 
that are shuttled around in two-dimensional microstructures to process and 
store quantum states at various locations. 

An initial systematic study showed that coherent transport of ions in linear 
trap arrays is possible with nearly no loss in contrast during the motion 
\cite{rowe}. In this experiment an adiabatic transport of a qubit was 
performed over a distance of $1.2~mm$ in a time span of about $54~\mu s$ with 
negligible heating. Currently, there are strong efforts under way to 
demonstrate the possibility of building large-scale ion trap structures. 
For example, suggestions have been made to combine 
miniaturized ion chips directly with CMOS electronics to handle the resources 
required to control the many electric potentials \cite{kimslusher}. Moreover, 
fast transport requires excellent experimental control of all these potentials.

A detailed scheme of how a viable architechture of an ion trap processor could 
look has been recently studied by Steane \cite{steane}, fully incorporating 
quantum error correcting codes. The physical gate rate of this proposed 
300 qubit processor unit was found to be limited by
\begin{align}
 \tau_g = \frac{2}{\nu_\text{COM}}+\frac{10}{\nu_\text{r}}+\tau_\text{cool}+
 \tau_p  \sim \text{a few}\ \mu s,
\end{align}
with the first two terms being an average time of the part of a typical gate 
that involve motion that is times for splitting ($\sim2\nu_\text{COM}^{-1}$), 
recombining and moving ($\sim 10\nu_\text{r}^{-1}$) a small ion string, where
$\nu_\text{COM}$ and $\nu_\text{r}$ are typical 
axial and radial trapping frequencies, respectively. 
The last two terms correspond to cooling after the transport has 
been done, and the time duration of conducting the actual phase gate, 
respectively. On the other hand, if large amounts of energy are transferred to 
the ions, longer cooling times might be needed. Inserting typical operating 
conditions shows that the first two terms make up 
a considerable part of the performance of the physical gate rate. In order to 
keep this part as small as possible we need designs for electrode structures 
enabling fast qubit transport.

In the following we present a theoretical framework that governs the transport 
dynamics of ions trapped in a time varying external potential. In Sect. 2 the 
equations of motion for the transport are derived. Sect. 3 discusses the 
general classical solution in terms of an Ermakov parametrization. This 
approach 
is useful to express the quantum approach presented in Sect. 5, which uses the 
Heisenberg picture  following the approach of Kim et al. \cite{kim}. 
In Sect. 4 
we point out some well-known properties of a quantum harmonic oscillator 
exposed to a transporting force for the simpler case when its frequency is 
kept constant. Sect. 5 presents the general quantum solutions and the 
interrelation between classical and quantum transport. Based on this framework 
we discuss in Sect. 6 a well-controlled regime for the transport and also 
include first order perturbations to the transport dynamics. In Sect. 7, we 
present numerical optimization routines to extract optimum switching of 
potentials for the transport and study miniaturization of electrode structures 
to estimate the required resources for a well-controlled transport. Finally, a 
simple electrode model is used to find a practical rule for the segmentation 
of ion traps revealing insight into the resources needed for large-scale 
layouts, that should be also applicable for more general trap arrays.


\section{Classical equations of motion}
A linear segmented Paul trap, e.g. as  used in recent experiments \cite{Cat, 
teleportation, QEC,QFT, stick, hensinger}, consists typically of two alumina 
wafers with 
gold coated electrode surfaces of a few micrometer thickness. The slotted 
wafers provide electrical RF and DC fields for 3D confinement of ions. The 
arrangement for control electrodes is schematically sketched in 
Fig.\ref{railwaytrack}  where only a single layer is shown. The confinement 
along the $x$-axis is achieved solely by electrostatic fields whereas the 
remaining two orthogonal radial directions correspond to a dynamical trapping 
by ponderomotive RF forces. In this article we limit ourselves to transport 
along a single dimension $x$ from $-b/2$ to $b/2$. If we denote the coordinate 
of the ion in the laboratory frame by $q$ then we have from Newton's equation 
of motion
	\begin{align}\label{ClassicalEqOfMotion}
	 	\ddot{q}(t)+\frac{Q}{m} \pd{\phi(q,t)}{q}=0,	
	 	   \qquad  q(-t_0)=-\frac{b}{2}, \quad \dot{q}(-t_0)=0,
	\end{align}
with two initial conditions as the equations on the rhs; 
$Q$ is the elementary charge and $m$ the mass of the transported ion. 
We assume a time interval and location of the ion starting at $-t_0$ and 
$-b/2$, and ending at $+t_0$ and $b/2$, respectively. 
In order to make use of coherent states of a harmonic oscillator (that do not 
spread in time) we are interested in designing the time-dependent electrical 
potential as
	\begin{align}
	  \phi(q,t)\sim\phi_\text{local}(q-q_0(t))+\varphi_0(t),
	\end{align}
where $\phi_\text{local}(q)\sim m \omega_0^2 q^2/2 Q$ is purely quadratic with 
constant curvature in a sufficiently large range around the minimum, and 
$\varphi_0(t)$ is a time-dependent offset with no influence on the dynamics.
Here, we prescribe the dynamics by specifying a desired harmonic frequency 
$\omega_0$ and the temporal shift of the harmonic well by a {\it transport 
function} $q_0(t)$. The residual, uncontrolled force caused by insufficient 
flexibility in creating the desired harmonic potential deteriorates the 
transport performance. Its effect can be described by the difference potential 
or residual acceleration, i.e.
	\begin{equation}\label{defphires}
	  \phi_\text{res}(q,t)=\phi(q,t)-\varphi_0(t)-
	    \frac{m \omega_0^2 (q- q_0(t))^2}{2 Q} \qquad\text{and}\qquad
	   a_\text{res}(q) = -\frac{Q}{m}\pd{\phi_\text{res}(q)}{q},
  \end{equation}
respectively.  Due to imperfect realization of the harmonic well 
$\phi_\text{res}(q,t)$ adds fluctuating parts to the ideal harmonic potential 
as a function of position or time, critically depending on the electrode 
structure used. In section 7 we will discuss a numerical scheme for 
approximating $\phi(q,t)$ based on superpositions of individual electrode 
potentials in an optimal way. 

We finally can write down the classical equation of motion 
	\begin{equation} \label{ClassicalEqOfMotion2}
	  \ddot{u}+\omega_0^2 u  = -\ddot{q}_0(t)+  a_\text{res}(u+q_0(t))
	\end{equation}
which we transformed into a frame moving with $q_0(t)$ by $u=q-q_0$. The net 
acceleration on the rhs corresponds to an external force and displaces the ion 
from its equilibrium position $u=0$ in this frame. Since we will treat only 
the first two perturbation terms we expand the final equation of motion 
around the minimum of the well and rearrange some terms to get
	\begin{equation}\label{finaleqofmotion}
	  \ddot{u} + \omega_0^2 \Bigl(1-a'_\text{res}[q_0(t)]/\omega_0^2 
	     \Bigr) u- a''_\text{res}[q_0(t)]/2~u^2 
	  		+ \ldots =  
	  		- \ddot{q}_0(t)+a_\text{res}[q_0(t)]
  \end{equation}
 where primes denote differentiation with respect to $u$. 
For the following discussion we abbreviate 
$\omega^2(t)=\omega_0^2 (1-a'_\text{res}[q_0(t)]/\omega_0^2)$ and write 
$f(t)=- \ddot{q}_0(t)+a_\text{res}[q_0(t)]$ for the rhs of 
Eq.(\ref{finaleqofmotion}). For certain electrode structures, we can disregard 
terms involving the second and higher order derivatives of $a_\text{res}(q)$ 
(cf. Section 7). We will make this assumption throughout the paper. In that 
case Eq.(\ref{finaleqofmotion}) simplifies to the equation of motion of a 
parametrically driven and forced harmonic oscillator with the Hamiltonian 
\begin{align}\label{hamiltonian}
		 \calH(t) = \frac{p^2}{2 m}+ 
		   \frac{m\omega^2(t)}{2}u^2- m f(t) u
  \end{align}
and $p=m \dot{u}$.

\section{Classical dynamics of ion transport}
To obtain a general classical solution with an arbitrary frequency modulation 
we first consider the formalism which is most often used in conjunction with 
time-dependent invariants within so called Lewis-Riesenfeld methods 
\cite{lewis68}. These approaches have been shown to be successful in the 
quantization of time-dependent harmonic oscillators with many different kinds 
of time-dependencies. Here, we discuss the general classical solution using 
the Ermakov equation and its generalized phase equation for time-dependent 
frequencies. We then employ in section 5 the approach of Kim et al. \cite{kim} 
to express the general quantum solution in terms of its classical solution. 
\subsection{Homogeneous Solution} Neglecting higher order terms we find the 
homogeneous part of the solution of Eq.(\ref{finaleqofmotion}) by setting 
$f(t)=0$, thus solving
	\begin{align}\label{homogeneousequation}
	  \ddot{u}_c + \omega^2(t) u_c =0
  \end{align}
for an arbitrary time-dependent frequency $\omega(t)$. For this, it 
is most convenient to make the ansatz
 	\begin{align}\label{ansatz}
	  u_1=\rho(t) e^{i\mu(t)} \qquad u_2=\rho(t) e^{-i\mu(t)},
  \end{align}
introducing an amplitude function $\rho(t)$ and a phase function $\mu(t)$,
both real. 
Inserting Eq.(\ref{ansatz}) into Eq.(\ref{homogeneousequation}) and 
considering real and imaginary parts results in the two equations
	\begin{align}\label{rawermakov}
	  \ddot{\rho}-\rho\dot{\mu}^2 + \omega^2(t) \rho = 0, \qquad 
	   2\dot{\rho}\dot{\mu} + \rho \ddot{\mu} = 0.
  \end{align}
$\rho$ is an integrating factor for the second equation on 
the right so that we can write
	\begin{align}\label{phaseequation}
	  \qquad \rho^2 \dot{\mu} = 1,
	\end{align}
where we have chosen the integration constant as $1$.
The constant on the rhs of Eq.(\ref{phaseequation}) has the SI units 
$m^2~rad/s$ that should be taken into account at the end. 
If we substitute this back into the first equation of 
Eq.(\ref{rawermakov}) we obtain the Ermakov equation for the 
amplitude function $\rho(t)$ 
  \begin{align}\label{ermakovequation}
	   \ddot{\rho} + \omega^2(t) \rho =1/\rho^3.
  \end{align}
For periods of constant frequency $\omega=\omega_0$ the general solution 
is \footnote{The general solution of this equation is easily obtained by
first using $\dot{\rho}$ as an integrating factor with the integration 
constant $2\omega\cosh \delta$. This equation is immediately 
transformed to a harmonic oscillator by $x=\rho^2-\cosh\delta/\omega$.}
	\begin{align}\label{solermakovequation}
	  \rho(t)= \pm \omega_0^{-1/2} 
	 \sqrt{ \cosh \delta + \sinh\delta \sin(2\omega_0 t+\theta)},
  \end{align}
where $\delta, \theta$ are constants of integration, their values depend on 
the past evolution \cite{lewis68}. The solution for the generalized phase is 
easily obtained once $\rho$ is known. From Eq.(\ref{phaseequation}) we have
	\begin{align}\label{integratedphaseequation}
	  \mu(t)= \int^t_{-t_0} dt' \rho(t')^{-2}.
  \end{align}
The general homogeneous solution is then given by 
 	\begin{align}\label{homogeneoussolution}
	  u_h(t)= a_c \rho(t) \cos (\mu(t)+\varphi),
  \end{align}
with the classical amplitude $a_c$ and initial phase  $\varphi$ fixed by the 
initial conditions.
 
\subsection{Green's function and general solution}
We use the general framework of Green's functions to define a particular 
solution to the inhomogeneous case of Eq.(\ref{finaleqofmotion}), where we 
again terminate the expansion, i.e. $a_\text{res}^{(n)}[q]=0$ for $n\ge 2$, to 
stay in a harmonic regime.  Using the two independent homogeneous solutions of 
Eq.(\ref{ansatz}) we can determine the causal Green's function  
	\begin{align}
	  G(t,t')= \theta(t-t') \rho(t)\rho(t')\sin(\mu(t)-\mu(t')),
  \end{align}
with $\theta(t-t')$  the Heaviside function. Employing $G(t,t')$, a particular 
solution is given by
	\begin{equation}
	  u_{p}(t)= \int_{-t_0}^{t} dt' G(t,t') f(t')
	  =   \rho(t) \int_{-t_0}^{t} dt' \sin(\mu(t)-\mu(t')) \rho(t') f(t') .
  \end{equation}
For later convenience we define the auxiliary function
 $\zeta(t)=  i e^{-i\mu(t)} \int dt'  e^{i\mu(t')} \rho(t') f(t')$ 
 which will be useful for expressing the general quantum solution.
In this notation we can abbreviate the particular solution by 
$u_p(t)= \rho(t)\left\{\zeta(t)+\zeta^*(t)\right\}/2$.

Thus, we obtained the general solution as the sum of the general homogeneous 
solution Eq.(\ref{homogeneoussolution}) and a particular solution
	\begin{equation}\label{generalsolution}
	  u_c(t)= u_\text{h}(t) + u_{p}(t)\equiv \rho(t)/2 
	   \left\{a_c e^{i(\mu(t)+\varphi)}+\zeta(t)\right\}+c.c.\quad .
  \end{equation}
Higher derivatives, like velocity and acceleration, can easily be found from 
the general solution in Eq.(\ref{generalsolution}) by using the Leibniz rule. 
For initial conditions where we start in the classical ground state $a_c=0$ we 
define the quantity $\Xi(t)=\dot{u}_{p}(t)+i\omega(t) u_{p}(t)$, assuming that 
the transport starts at $-t_0>-\infty$, i.e. later than the infinite past, and 
demand that it takes a finite amount of time. With the help of the last 
definition the energy transferred to the oscillator at instants $t_1$ (where 
$f(t_1)\equiv 0$) is then given by
	\begin{align}\label{generalenergytransfer}
	  \calW(t_1)=m\left|\Xi(t_1)\right|^2/2,
  \end{align}
with
   \begin{equation}\label{gentransfampl}
		\hspace*{-5mm}
	\Xi(t_1) =   \int_{-t_0}^{t_1} dt' \rho(t')  \Bigl( \rho(t_1) 
	    \dot{\mu}(t_1) \cos(\Delta\mu_{1 t'}) + 
	     \{ \dot{\rho}(t_1)+i\omega(t_1)\rho(t_1) \}
	         \sin(\Delta\mu_{1t'}) \Bigr) f(t'),
  \end{equation}
and $\Delta\mu_{1t'}=\mu(t_1)-\mu(t')$. We will call $\Xi(t)$ the adiabatic 
suppression amplitude and its absolute square the adiabatic suppression 
factor. Thus, we have derived the classical energy transfer for arbitrary 
frequency evolutions and arbitrary external transport forces. To evaluate this 
expression one first must solve for the explicit time-dependence of $\rho$ 
and $\mu$ according to 
Eqs.(\ref{ermakovequation},\ref{integratedphaseequation})  by  integrating the 
Ermakov equation and the phase equation, and finally compute the transferred 
energy at different times using 
Eqs.(\ref{generalenergytransfer},\ref{gentransfampl}).

\subsection{Adiabatic limit}
Since we are mainly interested in an adiabatic solution we can simplify the 
last expression by considering adiabatic expansions of the homogeneous 
solution for a parametrically driven harmonic oscillator. We introduce an 
adiabatic time scale $\calT$ such that
	\begin{align}
	  \dot{\omega}/\omega \sim \calT^{-1}\quad \text{for}\quad 
	      \calT \gg \omega^{-1}.
  \end{align}
The general adiabatic expansion of the differential equations 
Eqs.(\ref{ermakovequation},\ref{integratedphaseequation}) is readily obtained 
\cite{kulsrud}
	\begin{align}\label{adiabaticexpansion}
   \rho(t)\sim \frac{1}{\sqrt{\omega(t)}}+\frac{1}{8}
     \frac{\ddot{\omega}(t)}{\omega(t)^{7/2}}-\frac{3}{16}
      \frac{\dot{\omega}(t)^2}{\omega(t)^{9/2}}
   +\cdots \quad\text{and}\quad \dot{\mu}(t)\sim \omega(t)
   -\frac{1}{4}\frac{\ddot{\omega}(t)}{\omega(t)^2}+\frac{3}{8}
   \frac{\dot{\omega}(t)^2}{\omega(t)^3}
	   +\cdots 
  \end{align}
This procedure is equivalent to a perturbative approach on the first term in 
Eq.(\ref{ermakovequation}) \cite{lewis68}. We require that at instants $-t_0, 
t_1$, i.e. at times when we measure the oscillator's energy, the frequency has 
settled into a constant. Also for the following discussion we define that the 
oscillator's initial frequency at $-t_0$ is $\omega(-t_0)=\omega_0$, so that 
$\rho(-t_0)=1/\sqrt{\omega_0}$, $\dot{\rho}(-t_0)=0$. Taking into account only 
the lowest order of the expansion in Eq.(\ref{adiabaticexpansion}) the 
expression in Eq.(\ref{gentransfampl}) reduces to
	\begin{align}\label{firstorderheating}
		\Xi(t_1) = \sqrt{\omega(t_1)}
	  \int_{-t_0}^{t_1} dt' f(t')\omega(t')^{-1/2}
	  e^{i\Delta\mu_{1 t'}}  \qquad
	  &\text{with}\quad \Delta\mu_{1 t'}=\int_{t'}^{t_1} d\tau\ 
	  \omega(\tau)
  \end{align}
providing the adiabatic energy transfer in the first order of 
frequency modulation.


\section{Quantum and classical, dragged harmonic oscillators with 
         constant frequency}
Husimi \cite{husimi} and Kerner \cite{kerner} independently considered the 
forced quantum mechanical oscillator and found exact analytical expressions 
for their wavefunctions and propagators. We review some of their early ideas 
because they provide insight into the close relationship of the quantum and 
classical solution. In this paragraph we assume the frequency is independent 
of time. The corresponding Hamiltonian is given by Eq.(\ref{hamiltonian}) with 
$\omega(t)=\omega_0$.

Following Husimi and Kerner, we can "uncouple" the classical oscillation  
by the transformation
\begin{align}\label{phasetransf}
 \Psi(u,t)=\phi(u',t)\exp(i m \dot{u}_c u'/\hbar),
\end{align}
with $u'=u-u_c$ and $u_c$ at first undefined. Inserting 
Eq.(\ref{phasetransf}) into the time-dependent Schr\"odinger equation gives
 \begin{equation}\label{schroedinger}
   i\hbar\pd{\phi}{t}=\left( -\frac{\hbar^2}{2m}\pd{^2}{u'^2}
   +\frac{1}{2}m\omega_0^2 u'^2 \right) \phi
 	+ m(\ddot{u}_c+\omega_0^2 u_c -  f)u' \phi 
 	-(m/2)(\dot{u}_c^2-\omega_0^2u_c^2+2 f u_c)\phi.
 \end{equation}
On the rhs we see that we can make the second term vanishing if we choose 
$u_c$ to satisfy
\begin{align*}
 \ddot{u}_c+\omega_0^2u_c-f=0,
\end{align*}
i.e. if $u_c$ satisfies the classical solution of Eq.(\ref{hamiltonian}).
With this choice one can easily identify the classical action  
$L(t)=(m/2)(\dot{u}_c^2-\omega^2u_c^2+2 f u_c)$ 
of a forced harmonic oscillator in the third term on the rhs of 
Eq.(\ref{schroedinger}). Furthermore, if we make the ansatz
\begin{align*}
  \phi(u',t)=\chi(u',t) \exp\left[{\frac{i}{\hbar}
  \int_{-\infty}^t dt' L(t')}\right],
\end{align*}
we can absorb this term as a time-dependent phase into $\phi$. 
The remaining part of the wavefunction, $\chi$, then  needs only to 
obey the usual harmonic oscillator wave equation in the frame defined 
by the classical trajectory with its internal coordinate $u'$ 
\begin{align}\label{HOSchroedinger}
  i\hbar\pd{\chi}{t}=\left( -\frac{\hbar^2}{2m}\pd{^2}{u'^2}
   +\frac{1}{2}m\omega_0^2 u'^2 \right)\chi.
\end{align}
In this way one can achieve a separation of the forced harmonic oscillator 
from the unforced oscillator in a frame moving with the classial trajectory. 
The wavepacket does not become deformed by the homogeneously acting force. The 
quantum solution becomes displaced and only a phase is accumulated.

To determine further properties we can assume now a stationary state with 
energy $\epsilon_n$ for the solution of Eq.(\ref{HOSchroedinger})
\begin{align*}
  \chi_n(u',t)=u_n(u')\exp(-i \epsilon_n t)\qquad\epsilon_n=(n+\tfrac{1}{2})
   \hbar\omega_0,
\end{align*}
and evaluate transition probabilities at time $t$ 
for the oscillator to be in the number state $u_m$ if it was initially in 
the number state $u_n$    
\begin{align*}
  P_{mn}(t)=\left|\int_{-\infty}^\infty u_m(u-u_c(t)) u_n(u) 
  e^{i m \dot{u}_c(t) u /\hbar}du\right|^2.
\end{align*}
Husimi and Kerner showed that these transition moments can be 
evaluated analytically 
\begin{equation}\label{transprob}
   P_{mn}(t)=(\mu!/\nu!) \gamma^{\nu-\mu} e^{-\gamma} 
   (L_\mu^{\nu-\mu}(\gamma))^2
  \qquad\text{with}\quad \gamma(t)=m/2\hbar\omega_0~ |\dot{u}_c+i\omega u_c|^2
\end{equation}
by using generating functions for the Hermite polynomials 
\cite{husimi,kerner}. In Eq.(\ref{transprob}), $\nu$ is the greater while 
$\mu$ is the lesser of $m$ and $n$, respectively. 
$L_\mu^{\nu-\mu}$ denote the associated 
Laguerre polynomials, and its time-dependent argument $\gamma(t)$ describes 
the classical energy transfer in units of $\hbar\omega_0$. From 
Eq.(\ref{transprob}) we see the classical character of the quantum solution: 
the transition probabilities are solely defined by the classical quantity 
$\gamma(t)$. Also, if we consider starting from the ground state $n=\mu=0$ and 
using $L_0^{\nu}(\gamma)\equiv 1$ the probability distribution $P_{m0}$ 
becomes a poissonian, and thus we find the signature of a coherent state.

With this relation the expectation values for the mean energy and the 
dispersion of the energy distribution are then immediately obtained
\begin{align}\label{encompQM}
	\langle E_m\rangle_n &\equiv \hbar\omega_0 \Big( \sum_m m\ 
	P_{mn}+1/2\Big) = \hbar\omega_0 \Big( n+\gamma+1/2\Big) =
	\epsilon_n + \hbar\omega_0 \gamma  \\ \nonumber
	\langle (\Delta E_m)^2 \rangle_n &\equiv (\hbar\omega_0)^2 
	 \left\langle (m-\langle m\rangle)^2 \right\rangle= 
   (\hbar\omega_0)^2 (2 n+1) \gamma=2\epsilon_n\ \hbar\omega_0 \gamma\ ,  
\end{align}
where $\epsilon_n$ is the initial energy before the force acts on the 
wavepacket. This is indicated in Eq.(\ref{encompQM}) by the subscripts on the 
lhs. Corresponding expressions for the classical solution
\begin{align}\label{encompCl}
	\langle E\rangle_{E_0}  = E_0 + \calW  \qquad 	
	\langle (\Delta E)^2 \rangle_{E_0} = 2  E_0  \calW
\end{align}
are found if we average over the initial classical phase that are completely 
analogous to the quantum solutions.\footnote{This result is easily derived by 
averaging the general classical solution in Eq.(\ref{generalsolution}) given 
in Sect. 4 for a constant frequency over the phase interval $[0, 2\pi]$.} 
$E_0$ is the classical energy before the transport and 
$\calW\equiv\hbar\omega_0\gamma(t)$ the classical energy transfer. The mean 
energy and the energy spread increase linearly with the energy transfer in 
both solutions although the energy distributions of the classical and quantum 
solution are quite different \cite{husimi}. Also, the zero point energy makes 
a difference between the classical and quantum description. If the system is 
initialized in its quantum ground state, transport can create a dispersion of 
the wavepacket due to $\epsilon_0>0$, while this is not the case for the 
classical ground state, i.e. if $E_0=0$.


\section{The dragged quantum harmonic oscillator} 
Many methods have been developed to find exact quantum states of time-
dependent oscillators. The generalized invariant method by Lewis and 
Riesenfeld \cite{lewis68} has been very successful in finding exact quantum 
motion in terms of wavefunctions and propagators. For the interpretation of 
time-dependent quantum systems and for showing its relationships to their 
classical solution, however, the Heisenberg picture is more appropriate since 
the Heisenberg operators for position and momentum obey similar equations of 
motion than the corresponding classical quantities. In this paragraph we aim 
to interpret the quantum solution using its classical analogue and therefore 
use the general approach of Kim et al. \cite{kim} that is based on the general 
invariant theory but acts in a Heisenberg picture, in contrast to the original 
approach.

The general invariant theory starts out by defining an invariant operator 
$I(t)$ that satisfies the Heisenberg equation of motion. Ji et al. \cite{ji} 
used a Lie algebra approach to find the most general form of the solution with 
some integration constants $c_i,~i=1,2,3,$ arbitrary defining the initial 
conditions (see discussion at the end of this paragraph). If we fix these 
parameters according to the conditions of Eq.(3.4) in \cite{jisqueezed} the 
generalized invariant is of the form
\begin{align}\label{geninvariant}
 I_T(t)&=\omega_I\left(B^\dagger(t)B(t)+\frac{1}{2}\right),
\end{align}
with $\omega_I$ as a constant of motion, and the annihilation and 
creation operators are
\begin{align}\label{creatannihoperators}
  B(t)=\sqrt{\frac{m}{2}}\Big\{\left(\rho^{-1}-i\dot{\rho}\right)\hat{q}(t) 
  -\zeta\Big\} +i\frac{\rho}{\sqrt{2m}}\hat{p}(t) \nonumber\\
  B^\dagger(t)=\sqrt{\frac{m}{2}}\Big\{\left(\rho^{-1}+i\dot{\rho}\right)
  \hat{q}(t) -\zeta^*\Big\} -i\frac{\rho}{\sqrt{2m}}\hat{p}(t).
\end{align} 
It holds that $[B(t),B^\dagger(t)]=1$, where $B(t),B^\dagger(t)$ are solely 
represented by the classical quantities $\rho=\rho(t), 
\mu=\mu(t),\zeta=\zeta(t)$ as introduced in previous paragraphs. 
$\hat{q}(t),\hat{p}(t)$ refer here to the Heisenberg operators for position 
and momentum and we have assumed in addition that $f(-t_0)=0$.

From the Heisenberg equations of motion for $B$, i.e. $d B(t)/dt=-
i[B(t),\calH(t)]$, one can obtain the simple time evolution for these 
annihilation and creation operators
\begin{align}\label{timeevolcreatannihop}
  B(t)\equiv e^{-i \mu(t)}B(-t_0)  \qquad   B^\dagger(t)\equiv 
  e^{i \mu(t)}B^\dagger(-t_0),
\end{align}
with $\mu(t)$ the phase function.
Their evolution in time is a simple time-dependent phase-shift mediated by the 
generalized classical phase referenced to the initial time $-t_0$. This last 
property guarantees the time-independence of the invariant and the equivalence 
to the Hamiltonian (if $f(-t_0)=0$) at the time $-t_0$: 
\begin{align}\label{equivalence}
 I_T(t)=I_T(-t_0)=\calH(-t_0).
\end{align}
Following Kim et al. we can equate hermitian and anti-hermitian parts 
on both sides of Eqs.(\ref{timeevolcreatannihop}) by using
the relations Eqs.(\ref{creatannihoperators})
to determine the time-dependent Heisenberg 
operators for position and momentum 
\begin{align}\label{qop}
 \hat{q}(t)=\rho(t)\left\{\hat{q}(-t_0)\sqrt{\omega_0}\cos\mu(t)
  +\frac{\hat{p}(-t_0)}{m\sqrt{\omega_0}}\sin\mu(t)\right\}+u_p(t)
\end{align}
\begin{align}\label{pop}
 \hat{p}(t)=&\hat{q}(-t_0)m\sqrt{\omega_0} 
  \left[\dot{\rho}(t)\cos\mu(t)-\rho(t)^{-1}\sin\mu(t)\right]\nonumber \\
       &+\frac{\hat{p}(-t_0)}{\sqrt{\omega_0}}\left[\rho(t)^{-1}
       \cos\mu(t)+\dot{\rho}(t)\sin\mu(t)\right] 
       +m \dot{u}_p(t),
 \end{align}
where $\hat{q}(-t_0),\hat{p}(-t_0)$ denote position and momentum 
operator at time $-t_0$, respectively. 
Similar as $u_p(t)$, the classical velocity can be expressed as 
$2\dot{u}_p(t)=\left\{ (\dot{\rho}-i\rho^{-1})\zeta+ 
(\dot{\rho}+i\rho^{-1})\zeta^*\right\}$.

Our chosen initial conditions, Eq.(\ref{equivalence}), cast momentum 
and position operators into their standard form 
\begin{align}\label{opt0}
 \hat{q}(-t_0)=\frac{1}{\sqrt{2m \omega_0}}\left\{B+B^\dagger\right\}\qquad
 \hat{p}(-t_0)=-i \sqrt{\frac{m\omega_0}{2}}\left\{B-B^\dagger\right\}
\end{align}
taking $\hbar=1$.
Kim et al. define a more general Fock state space based on number states of 
the invariant rather than on the Fock state space of the Hamiltonian and point 
out its importance and advantegeous properties. However, due to our choice of 
the initial conditions these two state spaces are identical and their 
distinction is irrelevant for our discussion. We can define the Fock basis in 
the usual way by taking the operators at $-t_0$ according to
\begin{align}
 |n\rangle_B&=|n,-t_0\rangle_B \quad\text{with}\quad |n,t\rangle_B=
 \frac{B^{\dagger n}(t)}{\sqrt{n!}}|0,t\rangle_B, 
\end{align}
where the vacuum state $|0,t\rangle_B$ is extracted from $B(t)|0\rangle_B=0$.
Furthermore, we introduce the time-independent coherent states in this 
Fock basis
\begin{align}\label{coherentstate}
 |\alpha\rangle&=e^{-|\alpha|^2/2}\sum_{n=0}^{\infty}
 \frac{\alpha^n}{\sqrt{n!}} |n\rangle_B,
\end{align}
with the complex amplitude $\alpha=|\alpha|e^{-i\varphi}$, 
because these states are the closest quantum equivalent to the classical 
solution and include the oscillator ground state for $\alpha=0$.
With these definitions the expectation values for the Heisenberg position 
and momentum operators from Eqs.(\ref{qop},\ref{pop}) 
can be calculated using Eq.(\ref{opt0}) and Eq.(\ref{coherentstate}) 
\begin{align}
 \langle\alpha|\hat{q}(t)|\alpha\rangle &=\sqrt{\frac{2}{m}} \rho(t) 
 |\alpha| \cos [ \mu(t)+\varphi ]+u_p(t) \nonumber\\
 \langle\alpha|\hat{p}(t)|\alpha\rangle &=\sqrt{2 m}|\alpha|\Big\{
       \dot{\rho}(t)\cos [ \mu(t)+\varphi]
       -\rho^{-1}(t) \sin [ \mu(t)+\varphi ]
       \Big\}+m \dot{u}_p(t) \\ \nonumber
       &\equiv m \frac{d}{dt} \langle\alpha|\hat{q}(t)|\alpha\rangle.
\end{align}
This way we retrieve exactly the same form for the mean values of position and 
mometum for the quantum solution as we obtained in Eq.(\ref{generalsolution}) 
for the classical solution. If we disregard the zero point energy in 
$\langle\alpha| \calH(-t_0)|\alpha\rangle/\omega_0=|\alpha|^2+1/2\sim 
|\alpha|^2$ and set the matrix element equal to the potential energy at a 
classical turning point, we have $a_c\sim \sqrt{2}x_0 |\alpha|$ making the 
homogeneous solution of the classical and quantum formulations and hence the 
total solution identical. Here, $x_0=\sqrt{\hbar/\omega_0 m}$ is the extension 
of the ground state wave function of the harmonic oscillator. Alternatively, a 
full quantum description in the Schr\"odinger picture can be obtained by 
employing the time evolution operator that can be represented as a product of 
time-dependent displacement and squeezing operators \cite{kim,husimi}.

Similarly we can compute the dispersions of $\hat{q}(t), \hat{p}(t)$ in the 
coherent state
\begin{align}\label{dispersions}
 \langle \alpha|(\Delta q(t))^2|\alpha\rangle =  \rho^2/2m  \qquad\qquad\qquad
 \langle \alpha|(\Delta p(t))^2|\alpha\rangle = (\rho^{-2}+\dot{\rho}^2) m/2.
\end{align}
From the dispersion for the momentum we see that the wavepacket generally 
spreads solely due to the presence of the terms $\rho^2$ and $\dot{\rho}^2$. 
These matrix elements do not depend on the force because the force acts 
homogeneously in space and equally on the whole wavepacket.
After periods of frequency modulations  the dispersions in 
Eq.(\ref{dispersions}) 
are both time-dependent and exhibit oscillatory behaviour revealing a 
certain amount of squeezing \cite{ji}. For example, if we 
assume that after the transport we end up with a nonzero $\delta$ we can 
use the exact solution in Eq.(\ref{solermakovequation}) 
and evaluate the rhss of Eqs.(\ref{dispersions}).
Then, the dispersions for $q(t),p(t)$ are proportional to
\begin{align}
(\cosh\delta\pm\sinh\delta \sin (2\omega_0 t+\theta)),
\end{align}
distinguishable only by the $+$ and $-$ sign and constant prefactors, 
respectively. 
Therefore after the transport, the dispersions oscillate with twice the 
harmonic frequency and a relative phase shift of $\pi$. 
The strength of this squeezing oscillation is thus solely ruled by 
the classical quantity $\delta$.

Finally, our classical initial conditions $\rho(-t_0)=1/\sqrt{\omega_0}$ 
together with the choice of the free parameters $c_1=c_3=\omega_0/m, c_2=0$ 
which we used to define the annihilation and creation operators in 
Eqs.(\ref{qop},\ref{pop}), provide the correct initial dispersions of the 
quantum formulation in this approach 
\begin{align}
 \langle \alpha|(\Delta q(-t_0))^2|\alpha\rangle =   x_0^2/2  \qquad\qquad
 \langle \alpha|(\Delta p(-t_0))^2|\alpha\rangle =  \hbar^2 /2x_0^2.
\end{align}


\section{Transport dynamics in a well-controlled regime} 
\label{transportdynamics}
In the following we consider an idealized situation for the transport, i.e. we 
assume that we could produce arbitrarily shaped external potentials in the 
experiment while locally maintaining parabolic potentials around $q_0$, i.e. 
$1 \gg |a'_\text{res}(q_0)|/\omega_0^2$ and  $\vert \ddot{q}_0 \vert \gg \vert 
a_\text{res}(q_0)\vert$ for all positions $q_0$ or times  $q_0(t)$. Deviations 
from these ideal conditions due to constraints in realistic trap 
configurations will be evaluated in Sect. \ref{waveformdefinition}. In the 
ideal case we find from Eq.(\ref{firstorderheating})
	\begin{align}\label{simpleheat}
		\Xi(t_1) &=  -e^{i\omega_0 t_1}
	  \int_{-t_0}^{t_1} dt' \  e^{-i\omega_0 t'} \ddot{q}_0(t').
  \end{align}
For $t_0,t_1\rightarrow\infty$ we arrive at the well-known result that the 
transferred energy corresponds to the squared modulus of the Fourier transform 
of the time-dependent force at frequency $\omega_0$ \cite{landaulifshitz}. 
Since we can decompose any function into a sum of symmetric and anti-symmetric 
parts 
$q_0(t)=(q_0(t)+q_0(-t))/2~+~(q_0(t)-q_0(-t))/2\equiv q^{S}(t)+q^{A}(t)$ 
we can write
	\begin{align}\label{symmetry}
  |\Xi(t_1)|^2 =  \left|\int_{-t_0}^{t_1} dt' \ \sin(\omega_0 t') 
  \ddot{q}^A_0(t')\right|^2+
             \left|\int_{-t_0}^{t_1} dt' \ \cos(\omega_0 t') 
             \ddot{q}^S_0(t')\right|^2.
  \end{align}
The two parts increase the amount of the transferred energy independently. For 
a real transport, where start and stop positions differ from each other, we 
need anti-symmetric parts in the transport function. A simple conclusion from 
this is that any symmetric part of the transport function can only increase 
the transferred energy while not contributing to the purpose of the transport, 
therefore we only need to consider anti-symmetric functions as candidates for 
transport, i.e. we take $q^S_0(t)\equiv 0$. By partially integrating 
Eq.(\ref{simpleheat}) two times and using initial conditions for the start and 
stop position and velocities Eq.(\ref{ClassicalEqOfMotion}), we can also 
rewrite the integral in Eq.(\ref{simpleheat}) as a direct functional of 
$q_0(t)$ that has a similar appearance but with an additional term. By symbols 
in this text with an extra tilde we denote quantities that are divided by the 
half of the transport distance $b/2$, e.g. $\tilde{q}_0(\pm t_0)=q_0(\pm 
t_0)/(b/2)=\pm 1$. 

The average number of vibrational quanta transferred during transport can now 
be calculated from Eq.(\ref{generalenergytransfer}) 
\begin{equation} \label{normalizedEnTransf}
 \gamma(t_0)=m b^2\omega_0|\tilde{\Xi}(t_0)/\omega_0|^2/8\hbar.
\end{equation} 
The energy increase in a transport therefore scales quadratically with the 
transport distance if the time span is fixed. Before we systematically study 
expression Eq.(\ref{simpleheat}) we will consider two examples for which 
analytical solutions exist.

\subsection{Two analytical examples}
\begin{figure}
\includegraphics[width=65mm]{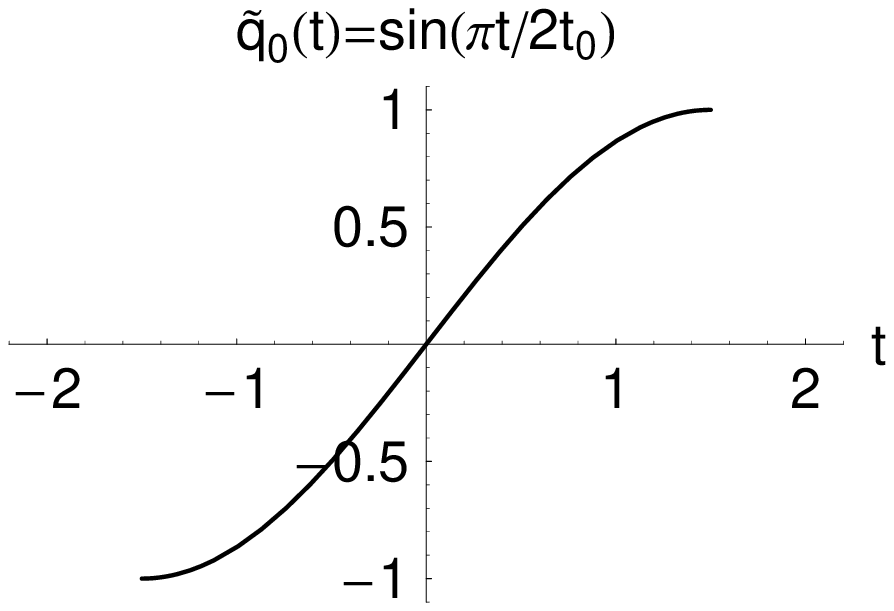}~a)
\hfil
\includegraphics[width=80mm]{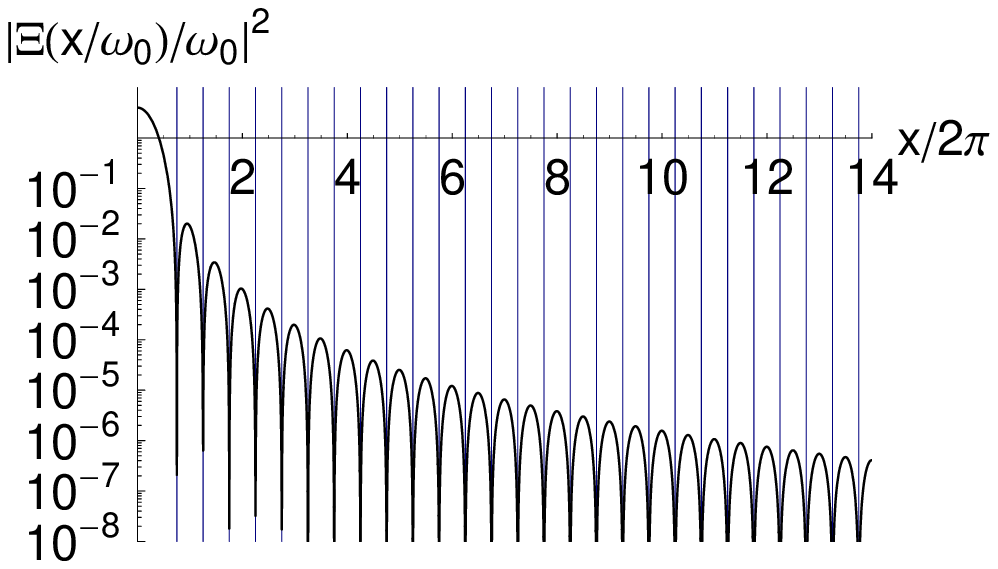}~b)
\caption{Sinusoidal transport. In \textbf{a)} a normalized transport 
   function for a sine transport is shown with a transport in 
   $2t_0=3$ time units.   Figure \textbf{b)} shows the energy transfer 
   as a function of $x/2\pi=t/T$, with $T=2\pi/\omega$ the period
   of the associated oscillation frequency. Zeros occur at the positions  
   $x/2\pi=(2n+3)/4$ with $n=0,1,\ldots$. See text for details.}
\label{sine}
\end{figure}

First we take a sine function for the transport function $q_0(t)$ as used 
in the experiments described in \cite{rowe}
\begin{align*}
 	\tilde{q}_0(t)=\sin(t \pi/2t_0) \qquad\text{for}\qquad -t_0<s<t_0.
\end{align*} 
A graph of this function is given in Fig.\ref{sine}a). Inserting 
$\tilde{q}_0(t)$ into Eq.(\ref{simpleheat}) we find
\begin{align}\label{sinh}
  \tilde{\Xi}(t_0)\equiv\tilde{\Xi}(x/\omega_0)=
       \omega_0 \frac{2 \cos(x)}{1-\left(2x/\pi\right)^2}\times 
       \text{phase} \quad\text{with}\quad x=\omega_0 t_0,
\end{align}
where we converted the time variables to dimensionless units, 
$x\equiv\omega_0 t_0$, so that $x/2\pi$ corresponds to the number of
oscillation cycles. In these variables, $|\Xi(t_0)/\omega_0|^2$ is 
independent of the frequency and plotted in Fig.\ref{sine}b). 
The energy transfer is decaying overall, but shows some oscillations arising 
from the dependence on the energy transfer on the phase of the internal 
oscillation at $t_0$. From Eq.(\ref{normalizedEnTransf}) we see that for an 
extreme nonadiabatic transport, i.e. $x\sim 0$, we have gained the full 
potential energy of $m\omega_0^2 b^2/2$. Depending on the exact transport 
duration we observe regular intervals where the energy drops to zero and no 
energy remains in the internal oscillator's motion after the interval length 
$2t_0$. This is due to the phase sensitivity of the transport. From 
Eq.(\ref{sinh}) we have the proportionality $|\tilde{\Xi}|^2\propto 
(1+\cos(2x))$, so that we expect the first zero for $x/2\pi=1/4$. 
However, for a transport in a harmonic well we need at least half an 
oscillation period for the ion to move to the other turning point, therefore 
we have instead $x/2\pi=3/4$ which is seen in Fig.\ref{sine}b) 
as the first root of the adiabatic factor. The denominator in Eq.(\ref{sinh}) 
cancels the first root. The adiabatic energy transfer corresponds to the 
envelope of this function and is given by $4/(1-\left(2x/\pi\right)^2)^2$. As 
we will see in the following the decay of the envelope can be sped up for 
different choices of the transport function.

Typically, we want to have $m b^2\omega_0 |\tilde{\Xi}/\omega_0 
|^2/8\hbar\lesssim 1$ in order to limit the maximum transferred energy to a 
few vibrational quanta. Let us consider some typical parameters for traps 
currently in use; we choose the axial frequency $\omega=2\pi\cdot 
3~\text{MHz}$, a typical average transport distance of about four traps 
(=control electrode widths), i.e. $b=400~\mu m$, and $m=9.01218~u$ equal to 
the mass of a Beryllium ion. Then the adiabatic suppression factor should obey
\begin{equation}\label{criterion}
  |\Xi/\omega_0|^2 < 2\cdot 10^{-8}. 
\end{equation} 
In Fig.\ref{sine}b) we have not plotted the whole range until this criterion 
is fulfilled. It is satisfied for about $x/2\pi \gtrsim 30$. Thus, for the 
given case adiabatic transport happens on a rather long time scale, i.e. 
durations of $2\cdot x/2\pi=2t_0/T=60$ cycles. The transport in the experiment 
\cite{rowe} which has used this transport function was performed over three 
times this distance requiring that $|\Xi/\omega_0|^2$ is lower by a factor of 
$9$ more. The adiabatic envelope has decayed to this value at about 
$x/2\pi\sim 52$ yielding a transport duration of $2t_0/T=104$ cycles. Using a 
sine transport function the experimentally measured limit was around 
$2t_0/T=157$ oscillation cycles (where $2t_0=54~\mu s$ and 
$\omega_0=2\pi~2.9~\text{MHz}$). This appears reasonable because the electrode 
array that was used in \cite{rowe} was rather sparse, thus not allowing 
for full control 
and maintaining the conditions assumed in this paragraph properly. Also, the 
envelope in this region is quite flat; so within the uncertainties of the 
experiment, the experimentally observed limit is in reasonable agreement 
with our estimation.

\begin{figure}
 \includegraphics[width=65mm]{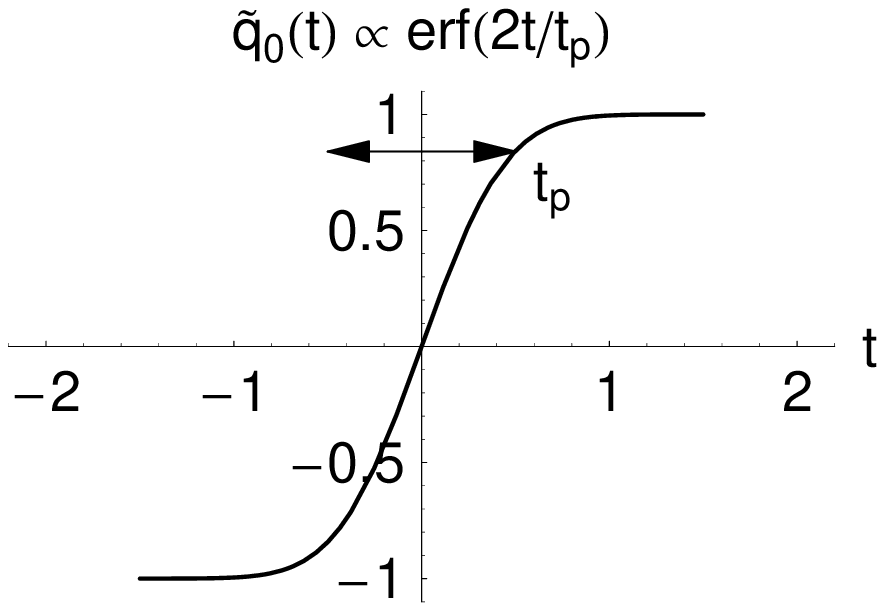}~a)
 \hfil
\includegraphics[width=80mm]{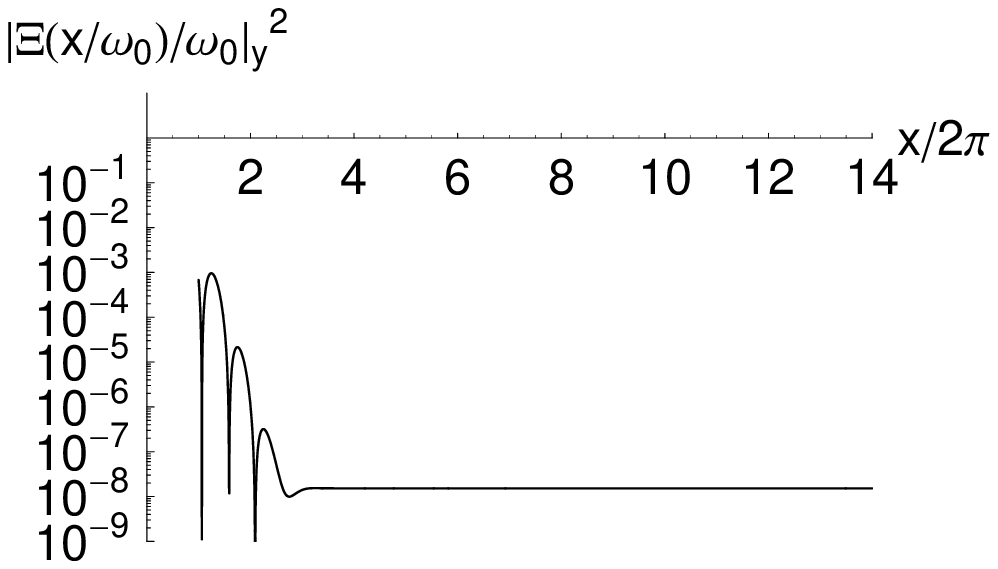}~b)
 \caption{Error function transport. \textbf{a)} graph of the renormalized 
 error function. \textbf{b)}
 Adiabatic suppression factor for the error function transport. Details are 
 given in the text.}
 \label{erf}
\end{figure}

We will look at  an error function transport which turns out to be 
advantageous to the sine function in the second example
\begin{align}\label{erftrapo}
  \tilde{q}_0(s)=\text{Erf}(2 s/t_p)/\text{Erf}(2 t_0/t_p)\qquad\text{for}
   \qquad -t_0<s<t_0,
\end{align}
where we renormalized it to arrive at the times $\pm t_0$ at the start and end 
position. In addition we have introduced another time $t_p$ which is nearly 
reciprocal to the slope of the transport function at the central point $t=0$. 
Fig.\ref{erf}a) is a graph of this function for $(t_p=1,t_0=3/2)$ in arbitrary 
time units.
Since we truncate the error function, we violate the second initial condition 
in Eq.(\ref{ClassicalEqOfMotion}) in a strict sense. However, we are 
interested only in settings  where $t_p/2t_0<1$, so that this constraint for 
the velocity can be satisfied arbitrarily well. The adiabatic suppression 
factor can be evaluated analytically 
\begin{align}\label{erflimit}
   \tilde{\Xi}(x/\omega_0)/\omega_0	&\sim 2e^{-y^2/16}\text{Re}
   \left\{ \text{Erf}\left[2x/y+i y/4\right]\right\}/ 
    \text{Erf}\left[2x/y\right] \quad\times \text{phase factor}\\ \nonumber
    &\stackrel{2x/y\rightarrow\infty}{\longrightarrow} 2 e^{-y^2/16}
    \quad\times\text{phase factor},
\end{align}
neglecting the part resulting from the finite initial and final velocities, 
and using the dimensionless variable $y=\omega_0 t_p$. Fig.\ref{erf}b) 
illustrates the situation for $y=12$ and $x/2\pi$ in a range of time intervals 
the same as for the sine transport but also satisfying $2t_0/t_p>1$. It is 
clear that by using the error function the transport can be performed much 
faster than with a sine transport function, while still satisfying inequality 
Eq.(\ref{criterion}). The full transport can now be performed in $2 x/2\pi\sim 
6$ cycles with tolerable energy transfer. Interestingly, taking the limit  for 
large $2x/y$ ratios in Eq.(\ref{erflimit}) removes the phase-sensitivity 
completely. However, we also want to note that the differences observed in 
these examples depend on experimental circumstances, e.g. for very short 
transport distances, the adiabatic suppression factor does not have to be 
small. In this case the differences between the adiabatic suppression factors 
is marginal in a qualitative sense. This can be seen in comparing 
Fig.\ref{sine}b) and Fig.\ref{erf}b) for cases when only about 
$|\Xi/\omega_0|^2<10^{-3}$ is required, e.g. occuring for transport distances 
much less than an electrode width. On the other hand we find an interesting 
and advantageous distance scaling behaviour from Eq.(\ref{erflimit}): 
transporting longer distances does not require much longer time intervals. For 
example, Steane \cite{steane} estimated that within large-scale operation for 
the processing of a typical gate an average transport distance of $\sim 22$ 
traps is needed. By employing an error function transport we find that this is 
feasible with less than a quantum of transferred energy using the parameters 
$(y=13,x/2\pi=4)$, i.e. in already about 8 oscillation cycles, only about a 
third more time than for a transport over $4$ traps. The average velocity for 
such a transport is then considerably higher.

\subsection{Near-optimum transport functions}
In an attempt to optimize the transport function we can expand expression 
Eq.(\ref{erflimit}) up to the first 
order correction \cite{rocco}
		\begin{align}\label{Xiexpansion}
    	\tilde{\Xi}(x/\omega_0)_y \stackrel{r\rightarrow\infty}{\sim}
    	2\omega_0 \left(e^{-b^2} -  \frac{e^{-r^2}}{\sqrt{\pi}\sqrt{r^2+b^2}}\ 
       \cos(2r b+\arctan(b/r))\right)/\text{Erf}\left[r\right]
       ~\times\text{phase factor},
		\end{align}
with $r=2 x/y\equiv 2 t_0/t_p$, $b=y/4$. Because of $2r b=x$ the zeros of the 
suppression factor $|\Xi|^2$ are equally spaced as in the sinusoidal example 
if we disregard the phase in the cosine function, i.e. half periods 
$\Delta(x/2\pi)=1/2$ since the particle can arrive from two different turning 
points at the end of  the transport. From this expansion it is clear that the 
ratio $r$ basically determines the magnitude of the second term on the rhs of 
Eq.(\ref{Xiexpansion}) and suppresses the phase sensitivity as it increases. 
If $r$ is chosen large enough the energy transfer is dominated by $b$. To find 
some conditions that are close to optimum we proceed in the following way: 
first, from the transport distance and achievable frequency we can evaluate 
the upper bound for the adiabatic suppression factor as in 
Eq.(\ref{criterion}). Because we have to satisfy $2t_0>t_p$ we then choose $b$ 
large enough to suppress the first exponential factor to fulfill the given 
criterion. This procedure defines the asymptotic value of energy transfer for 
large $x/2\pi$ as shown in Fig.\ref{erf}b). We then choose the interval length 
$x/2\pi=r b/\pi$ by defining $r$ such that we are just in the asymptotic 
range. The near phase-insensitivity can then be thought of as a result of the 
extremely slow start where the phase information in the limiting case 
$r\rightarrow\infty$ in Eq.(\ref{erflimit}) gets totally lost.

\subsection{High-frequency limit, adiabatic transport, and 
approximate trajectories}
To better understand the behaviour discussed in previous sections, we present 
a few more general considerations. We write the time-dependence of the 
transport function according to $q_0(t)=\vartheta_0(t/t_0)=\vartheta_0(\tau)$ 
so that $\vartheta_0$ only depends on the dimensionless variable $\tau$ (for 
the error function example we also keep the ratio $t_p/t_0$ fixed). Making the 
substitution $t'\rightarrow \tau$ in the integral  in Eq.(\ref{simpleheat}) we 
find
	\begin{align}
		\Xi(t_0)/\omega_0 &=  (\omega_0 t_0)^{-1}
	  \int_{-1}^{1} d\tau \  e^{-i\omega_0 t_0 \cdot\tau} 
	  \vartheta''_0(\tau) \quad\times\text{phase factor}\\
	     \nonumber
		    &\stackrel{\omega_0 t_0\rightarrow\infty}{\sim} 
		     2\cdot\sum_{n=0}^N 
    \frac{\vartheta_0^{(n+2)}(-1)}{(-\omega_0 t_0)^{n+2}}
    ~\cos(\omega_0 t_0+n \pi/2)
		    \quad\times\text{phase factor}.
	  \end{align}
The exponent in the integral relates the two time scales  in $\omega_0$ and 
$t_0$. Using the method outlined in appendix \ref{integralexpansion} we expand 
it into the sum given in the second line in the limit $\omega_0 t_0 
\rightarrow \infty$ assuming that $\vartheta_0$ is sufficiently smooth. In 
this expansion the derivatives at the start position (and end position due to 
anti-symmetry) define the energy transfer in the transport, and thus fully 
characterize the transport function for the transferred energy in the 
adiabatic limit. This provides us with a reason for the difference we observed 
above for the error function and sine examples. The second derivative  for the 
sinusoidal transport is nonzero at $\pm t_0$ and much larger than in the case 
of the error function. In the latter all derivatives are damped by a gaussian 
while the ones for the sine transport alternate. Furthermore, we see that we 
can in general decrease the transferred energy for larger values of the 
product $\omega_0 t_0$, i.e. by taking $\omega_0$ to inifinity (high-frequency 
limit), we can lower the adiabatic suppression arbitrarily, on the other hand, 
slowing down the motion by increasing the length of the duration of the 
transport $2t_0$, we move into the adiabatic regime. For infinitely slow 
motion we end up with zero transferred energy. These two limiting cases are 
formally equivalent because the energy transfer depends only on their relative 
time scale. We can perform the same expansion starting from 
Eq.(\ref{simpleheat}) directly and use the relation $u_c(t)=[\Xi(t)-
\Xi^*(t)]/2i\omega_0$ to find approximate trajectories valid in the same 
limits
	\begin{align}\nonumber
	  u_c(t)&= -\sum_{n=0}^N \frac{1}{\omega_0^{n+2}}
  	 \left[ \cos \left(\dfrac{n\pi}{2}\right) q^{(n+2)}_0(t)-
  	  \cos\left(\omega_0 (t+t_0)-\dfrac{n\pi}{2}\right) q^{(n+2)}_0(-t_0)
  	   \right].  	  	
	  \end{align}	  

\section{Regularized trap-electrode waveforms, potential fluctuations 
and aspect-ratio rule}\label{waveformdefinition}
\subsection{Determination of waveforms}\label{waveformequations}

So far, we have said nothing about how to determine the waveforms applied to 
the electrodes. As soon as we have the waveforms at hand for a given model 
electrode 
configuration we can determine the magnitudes of perturbations. This is done 
in the next section. Here, we seek optimum solutions for a given electrode 
structure in order to keep the uncontrolled part $\phi_\text{res}(q)$ of the 
total potential in Eq.(\ref{defphires}) small. The time-dependent electric 
potential is created by a linear superposition of the available control 
potentials $\phi_m(q)$ and dimensionless time-dependent amplitudes $a_m(t)$
of the form
	\begin{align}\label{suppotentials}
	  \phi(q,t)=\sum_m a_m(t) \phi_m(q).
	\end{align}
To optimize waveforms for the time-dependent amplitudes for the transport 
problem we find a measure of the discrepancy by integrating over the residual 
non-matched part according to
	\begin{align}\label{minimizationproblem}
	 \min_{a_m, \varphi_0}
	  \int_{q_0(t)-\delta q}^{q_0(t)+\delta q} \left|
	  \phi_\text{res}(q,t)\right|^2 dq \qquad \forall~t
	\end{align}
while $\phi(q,t)$ from Eq.(\ref{suppotentials}) enters here through 
Eq.(\ref{defphires}). For any time $t$ we want to find a set $a_m, \varphi_0$ 
for which expression Eq.(\ref{minimizationproblem}) is minimal. The 
integration is performed over an interval moving with the minimum of the 
parabolic potential well, i.e. $[q_0(t)-\delta q, q_0(t)+\delta q]$ and 
assuming a unity weight factor in the integrand. We do not consider in this 
range any lag of the ion due to acceleration and deceleration since for an 
adiabatic transport and experimental conditions the lag is much smaller 
compared to the optimization range. $\varphi_0(t)$ represents here another 
degree of freedom that does not perturb the dynamics but 
might allow one to more optimally choose 
the harmonic potential well by arbitrarily offsetting the desired 
parabolic potential for best fit. Condition Eq.(\ref{minimizationproblem}) is 
readily converted into a linear system of equations by taking partial 
derivatives for the amplitudes $a_m$ and $\varphi_0$, and setting them all 
equal to zero. The minimization problem in Eq.(\ref{minimizationproblem}) then 
reads
\begin{equation}\label{waveformequations2}
	\sum_{m=1}^{n_\text{el}} a_m(t) 
	    \underbrace{\int dq\ \phi_m \phi_j}_{{\bf S}_{a}}
	 + \varphi_0(t) 
	    \underbrace{\int dq\ \phi_j}_{\underline{S}_{\varphi_0}}
	 = \frac{m \omega^2}{2 Q} \underbrace{\int dq\   (q-q_0(t))^2   
	 \phi_j}_{\underline{K}} \quad  \forall~j,
\end{equation}
where we dropped the explicit integral bounds and arguments of the 
potentials $\phi_m(q)$ for the sake of simplicity.
Bold symbols denote matrices, underlined symbols vectors. 

The optimization problem can then be formulated in terms of the linear system
\begin{equation} \label{linearsystem}
{\bf S}_{+0}\cdot \underline{a}_{+0}\equiv
\left(\begin{array}{cc}
{\bf S}_{a} & \underline{S}_{\varphi_0}
\end{array}\right)\cdot
\left(\begin{array}{c}
\underline{a} \\
\varphi_0
\end{array}\right)= \eta 
 \underline{K} \quad  \forall  t,
\end{equation}
with $\eta=m \omega^2/2 Q$. All quantities are functionals of $q_0$ and for a 
given transport function $q_0(t)$, we need to solve the equations at every 
point in time. As a result we obtain the waveforms $a_m(t)\equiv a_m[q_0(t)]$. 

Typically we choose an optimization range of $2\delta q\sim 0.5 W$ for 
electrodes of width $W$. This is usually much smaller than the mean distance 
between most 
of the contributing electrodes to the center of the parabolic well. Thus, due 
to the  slow decay of the axial potentials, the curvatures of distant 
electrodes are similar, and their contribution differs locally only by a 
multiplication constant. This is particularly true for experimental situations 
where the high electrode density typically makes the system 
Eq.(\ref{linearsystem}) nearly singular. A straightforward least-square 
method, such as
\begin{equation}
  \underline{a}_{+0} = \text{arg}\min_{\underline{a}, \varphi_0} 
  \left\{\| {\bf S}_{+0}\cdot \underline{a}_{+0}- 
  \eta\underline{K} \|^2\right\},
\end{equation}
is therefore not well suited for finding waveform amplitudes. 
For high electrode 
density, a tiny step $q_0\rightarrow q_0+\delta$ might change the individual 
electrode amplitudes exponentially fast. In these cases the matrix ${\bf 
S}_{+0}$ in Eq.(\ref{linearsystem}) filters out too much information from 
$\underline{a}_{+0}$ to invert this system properly. In mathematical language 
these kind of problems belong to the family of {\it discrete ill-posed 
problems} that can be numerically solved  using regularization approaches 
\cite{hansen}. Here, the lost information is fed back in the minimization 
process via a Lagrangian multiplier concerning the smoothness of $a_m(t)$, or 
curvature etc. in amplitude space. If we apply a Tikhonov regularization to 
the given problem we have to solve
\begin{align} \label{regularization}
  \underline{a}_{+0,\nu} = \text{arg}\min_{\underline{a}, \varphi_0} 
  \left\{ \| {\bf S}_{+0}\cdot \underline{a}_{+0}- 
   \eta\underline{K} \|^2 + \nu^2 \| {\bf L} (\underline{a}_{+0}
   -\underline{a}_{+0}^*)\|^2\right\}
\end{align}
in order to determine smooth time-dependent waveform amplitudes $a_m(t)$.  In 
Eq.(\ref{regularization}) the regularization parameter $\nu$ corresponds to a 
weight factor between the original least-square minimization and the 
additional side constraints, while $\underline{a}_{+0}^*$ can be used to find 
solutions near a prescribed setting. The smoothing properties of this 
optimization originate from a common and simultaneous  minimization of both 
terms. $L$ is a linear operator that can be used to feed back different kinds 
of information to the amplitudes. For the results given here we took for $L$ 
the unity operator, and also $\underline{a}_{+0}^*=0$. Since we only want to 
limit the amplitudes $a_m$ to some appropriate experimental values and 
stabilize the solution, our interest is not to determine the overall minimum 
of Eq.(\ref{regularization}) in a self-consistent way. For our convenience we 
choose $\nu$ manually to make the parameters compatible with available 
technology.

We can summarize the advantages of these methods to the current optimization 
problem:
\begin{enumerate}
 \item The regularization method selects only nearby electrodes for creating 
       a local parabolic potential, and disregards
       tiny linear contributions from distant electrodes which would require 
       large amplitudes to effect small changes.
 \item The choice of the regularization parameter limits the amplitudes $a_m$ 
       to practical experimental values.
 \item It is robust against changing the electrode density (here, governed 
       by the widths W). This will be of importance in the next section.
 \item It stabilizes the output waveforms and smoothes sharp features in 
       the time-dependence of the amplitudes. 
       Different constraints can be set via the  $L$ operator, defining 
       bounds or curvatures in the amplitude space.
\end{enumerate}
For more detailed information we refer the reader to the mathematical literature 
\cite{hansen}. A typical example of a parabolic potential created through 
superposition of an array of electrodes and the time-dependence of amplitudes 
is shown in Fig.\ref{waves} and further discussed in the next section.


\subsection{Potential fluctuations and aspect-ratio rule}
\begin{figure}
\includegraphics[width=30mm,angle=90]{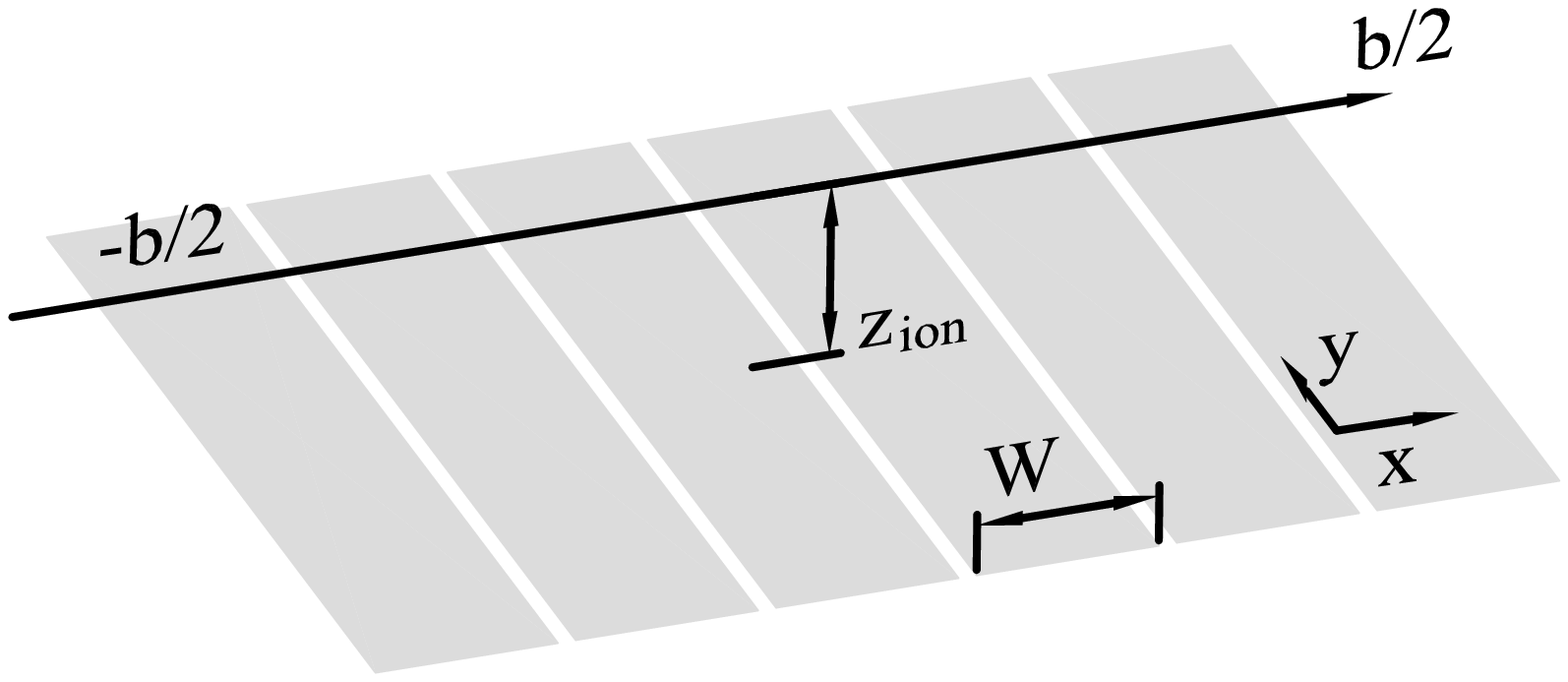}~a)
\hfil
\includegraphics[width=75mm]{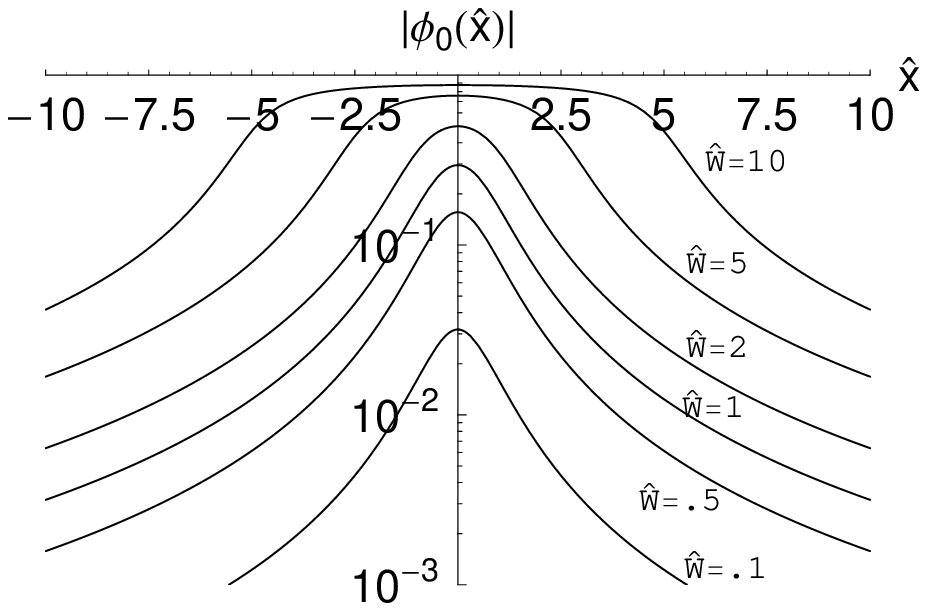}~b)
\caption{\label{railwaytrack} \textbf{a)} The 
control electrode array model discussed in the text with $x$ the axial 
coordinate, 
$z_\text{ion}$ the vertical distance of trapped ions to the electrode plane, 
and $W$ width of an electrode. No RF electrodes are shown to hold the ion
in radial directions. The electrode 'stripes' are assumed infinitely 
extended along $\pm y$, and arbitrarily many along $\pm x$. Ions are
transported along the long arrow from $-b/2$ to $+b/2$.
\textbf{b)} Modulus of the electric potential of a single electrode at height 
$z_\text{ion}$ along x (relative to the center of that electrode 
and normalized to $z_\text{ion}$). The electrode is embedded in an infinite 
ground plane and held at $-1~V$, and $\hat{W}=W/z_\text{ion}.$}
\end{figure}

Based on a reasonable multi-electrode structure we want to estimate how well 
we can meet the requirements on transport potentials stated above, in 
particular, how stringently we can meet $|a'_\text{res}(q_0)|/\omega_0^2 \ll 
1$ and  $\vert a_\text{res}(q_0)\vert \ll \vert \ddot{q}_0 \vert$. We employ 
the definition of waveforms and the method from the previous section for 
extracting the residual, uncontrolled potential $\phi_\text{res}$ from which 
we perturbatively derive the effect of imperfections on the transport. As a 
simple model electrode structure for transport in single and multi-layer 
traps, we use the ``railway track`` electrode configuration sketched in 
Fig.\ref{railwaytrack}a) which might be a simple model for transport in single 
and multi-layer traps \cite{kimslusher, stick, seidelin, home}. 
The transport occurs 
along the long arrow where  we assume the ion is held radially by RF fields and 
controlled axially  by the electrical fields arising from the potentials of 
the ``stripe`` electrodes depicted in Fig.\ref{railwaytrack}a). We are mainly 
interested in the scaling behaviour as a guideline for general design rules. 
Waveforms that are actually used in experiments should be based on more 
accurate numerical potentials and generalized versions of 
Eq.(\ref{waveformequations2}) for all three dimensions. We are finally 
interested in the trade off between adding electrodes, by shrinking the 
electrode distances/widths along x, and the amount of control  that is gained 
in that way.

\begin{figure}
\vspace*{0.2cm}
\includegraphics[width=55mm,angle=-90]{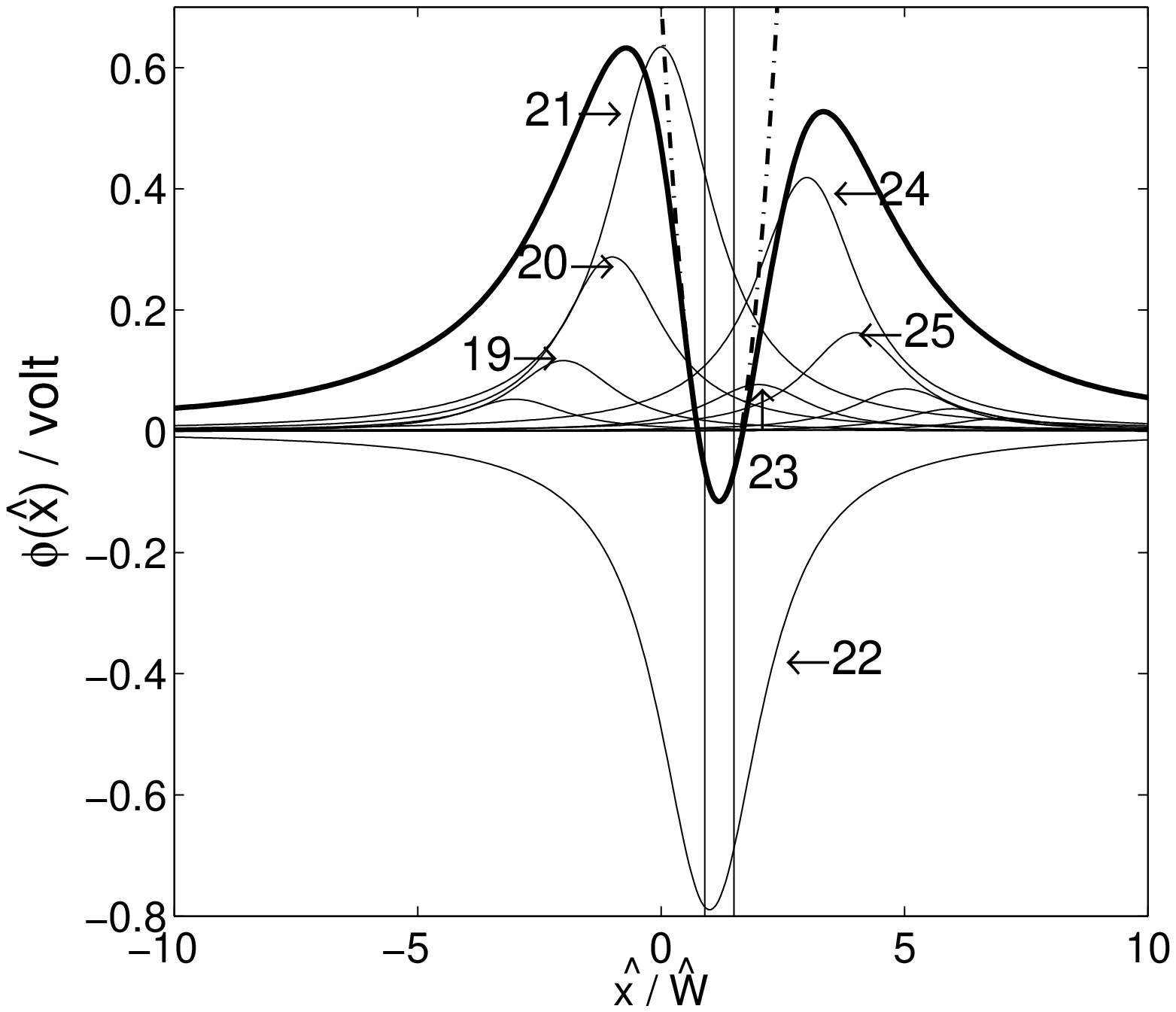}~a)\hspace*{1.3cm}
\includegraphics[height=65mm,angle=-90]{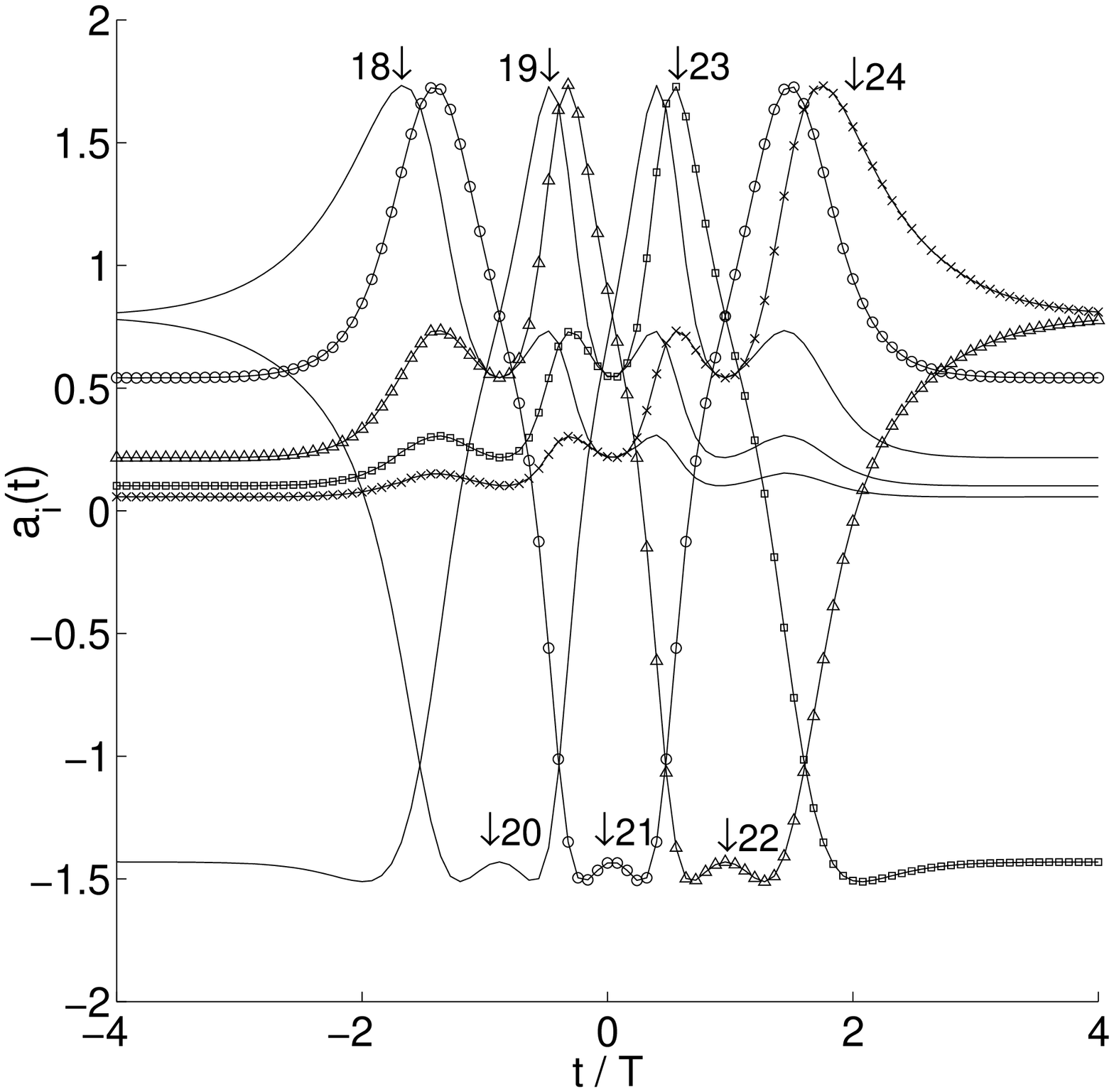}~b)
\caption{\label{waves}\textbf{a)} Creation of a harmonic well by superposing 
   potentials of the electrodes 
   of the stripe configuration of Fig.\ref{railwaytrack}a). 
   The fine lines represent individual electrode potentials that sum up to 
   the total potential represented by the thicker 
   line. The dashed-dotted line shows the ideal harmonic potential and the
   two vertical lines indicate the range of optimization at this given 
   location. By symmetry the extrema of the individual electrode potentials 
   are located at integer values of $\hat{x}/\hat{W}$, with the maximum of
   electrode potential 21 at the origin.   
   \textbf{b)} Waveform amplitudes of electrodes 18 to 24 for a transport 
   from trap 19 to 23 for $\hat{W}=1$. 
   Waveforms from more distant electrodes still contribute but are not shown 
   for the sake of clarity. 
   Amplitudes 21-24 are marked by symbols for better visualizing 
   their traces. The transport is done according to an error function,
   while the abscissa represents time in units of oscillation periods $t/T$.
}
\end{figure}

We can model this arrangement as a sum over the potentials $\phi_m(x)$ of 
several infinitely long (in the y-direction) stripe electrodes that are 
distributed along the x-axis
\begin{align}
	\phi(\hat{x}) = \sum_{m} a_m \phi_m(\hat{x})=
	    \frac{1 V}{\pi}\sum_{m} a_m\arctan 
	    \left( \frac{\hat{W}}{1+(\hat{x}-m \hat{W})^2-\hat{W}^2/4}\right),
\end{align}
where each $\phi_m(x)$ is the exact solution of the Poisson equation for an 
infinitely long stripe at position $m\cdot W$ that is embedded in a ground 
plane. 
For convenience we choose for the individual potentials in this basis set a 
potential on the electrodes of $1\text{Volt}=1~V$. We denote symbols with a hat 
as quantities normalized to the ion distance $z_\text{ion}$ to the surface, 
e.g. the normalized electrode width $\hat{W}=W/z_\text{ion}$. 
Fig.\ref{railwaytrack}b) shows the behaviour of $\phi_0(\hat{x})$ for various 
geometric aspect ratios $\hat{W}$. We see that a plateau-like structure starts 
to form for $\hat{W}\sim 2$ and larger resulting in small field gradients 
along the transport direction in the center of each stripe electrode. The 
maximum frequency at the center of electrode $m=0$ is obtained from
\begin{align} 
 \omega^2(\hat{W},z_\text{ion})=\left.
   \frac{Q}{m z_\text{ion}^2}\pd{^2}{\hat{x}^2}
   \phi_0(\hat{x})\right|_{\hat{x}=0}=-\frac{2 a_0}{\pi m}
   \frac{1 eV}{z_\text{ion}^2}\frac{\hat{W}}{(\hat{W}/2)^4+3\hat{W}^2/2+1}.
\end{align}
The proportionality $\omega\propto z_\text{ion}^{-1}$ is rather unusual and 
stems from the fact that we scale only a single dimension (along $x$). The 
last factor of the second equation is solely defined by the aspect ratio 
$\hat{W}$ and thus by the geometry of the trap. It exhibits a maximum for 
$\hat{W}\sim 0.78$ and decreases only significantly for small width-distance 
ratios $\hat{W}<0.5$. Using the mass of Beryllium, $\hat{W}=1$, amplitude 
$a_0=-2$ and $z_\text{ion}=40~\mu m$ as the ion surface distance for the 
surface-electrode trap as used in \cite{seidelin}, we find an axial frequency 
of about $\omega_0\sim 2\pi\cdot 9~\text{MHz}$. We chose this low value for $|a_0|$ 
to be compatible with typical maximum voltages as created by CMOS electronics 
\cite{kimslusher}. With these parameters we have created waveforms for the 
transport of a confining harmonic well utilizing the regularization approach 
of the last section. We used a set of $41$ electrodes while the transport was 
over four electrode widths, $\hat{b}=4$, around the central electrode $21$ of 
this array. For the transport we used the  error function Eq.(\ref{erftrapo}) 
of Sect.\ref{transportdynamics} with a transport duration of $8$ oscillation 
cycles $x/2\pi=4$ and $y=12$. We then determined the lowest order deviations 
from an ideal harmonic potential with constant trap frequency and 
controlled acceleration, $\omega(t)/\omega_0$ and 
$-a_\text{res}(q_0)/\omega_0^2$,
respectively, in Eq.(\ref{finaleqofmotion}) for the aspect-ratios 
$\hat{W}=(0.5, 1.0, 1.5, 2.0)$. Figs.\ref{waves}a) and b) illustrate an 
example for the superposed potentials, and for a set of waveform amplitudes 
for an error function transport in $8$ oscillation cycles.
   
The choice of the regularization parameter $\nu$ is not obvious, because we 
have to deal with a set of near-singular matrices all at once. As mentioned 
earlier we do not aim for self-consistent methods to determine $\nu$ and an 
absolute minimum of the expression Eq.(\ref{regularization})~\cite{hansen}. In 
our context we are more interested in a feasible implementation  compatible 
with given experimental constraints. The choice of the regularization 
parameter affects both the stability of the linear system and the size and 
smoothness of the amplitude vector $a_m$. In a strongly regularized inversion 
more stability is added to the solution, forcing the amplitudes to be of 
limited size. Because of this bound the solution can not closely approximate 
the desired shape of the potential anymore, so the deviations from the ideal 
case increase. A weak regularization scheme, on the other hand, adapts more 
closely to the desired potential shape, but reveals random fluctuations and 
noise on the solution waveforms $a_m(t)$. Also the singular behaviour 
increases dramatically with an increase of the number of electrodes, and 
larger parameters $\nu$ have to be chosen. This latter property makes a direct 
comparison of the results among various aspect-ratios $\hat{W}$ difficult. 
Nevertheless, we can make some qualitative and general statements.

\begin{figure}\hspace*{1.5cm}
\includegraphics[width=90mm,angle=-90]{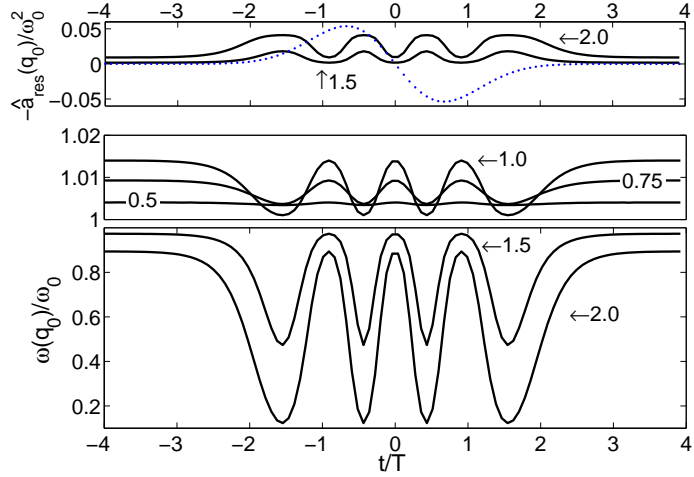}\vspace*{-1.7cm}
\caption{\label{waveforms} First and second order perturbation to the 
    transport (thick lines), and their dependence on 
    various aspect ratios $\hat{W}$. The abscissa $t/T$ corresponds to 
    the number of oscillation cycles with period $T=2\pi/\omega_0$.
    The labels on the graphs denote the value of $\hat{W}$.  
    (Upper panel)  Residual, uncontrolled acceleration 
    $-\hat{a}_\text{res}(q_0)/\omega_0^2$ (dimensionless) which
    should be compared to $\ddot{\hat{q}}_0/\omega_0^2$ (dotted line). 
    This is only given for the
    large aspect ratios $\hat{W}=1.5,2.0$. The graphs for smaller ratios 
    would be close to zero on this scale.
    (Middle and lower panels) Relative frequency modulation 
    $\omega(q_0)/\omega_0$ 
    during an error function transport for various
    aspect ratios. For explanations see text.}
\end{figure}

 The results for our sample configuration are summarized in 
 Fig.\ref{waveforms}. The upper graph in Fig.\ref{waveforms} displays the 
 uncontrolled acceleration $-a_\text{res}(q_0)/\omega_0^2$, and the middle 
 and lower 
 graphs the frequency modulation $\omega(t)$  for various aspect ratios. In 
 both figures a dramatic change of the curves is observed around $\hat{W}\sim 
 1$. While for smaller ratios the frequency fluctuations are in the percentage 
 range (middle panel), the  emulation of the potential for larger ratios is 
 much worse due to the constraint $|a_m|\le 2$ (lower panel). 
 Frequencies drop by more than 
 $90\%$ already for the $\hat{W}=2$ calculation. Only in the strongly 
 regularized scheme, did we find a direct correlation of the solution vector 
 $a_m$ to the choice of the regularization parameter. In the weakly 
 regularized scheme the amplitudes were limited by other lower bounds, and the 
 waveform solution appeared similar over a large range of $\nu$, 
 but exhibited a 
 much noisier behaviour. This enhanced sensitivity is an indication that 
 inclusion of more electrodes (smaller $\hat{W}$) does not improve the quality 
 of the solution anymore. The linear system becomes more singular and exhibits 
 more rank-deficiency, i.e. rows and columns become more equal and  their 
 inclusion adds more redundancy. For the given parameters we observed that for 
 $\hat{W}=0.5$ the transition from a regularized to a weakly regularized 
 solution occured. Therefore, our results indicate that $\hat{W}=0.5-1.0$ 
 should be optimal for the configuration discussed here. For larger aspect-
 ratios we found that the coverage of curvatures of the individual potentials 
 along the transport axis is not sufficient for the necessary amount of 
 control.

The other constraint, i.e. $\vert \hat{a}_\text{res}(q_0)\vert \ll \vert 
\ddot{\hat{q}}_0 \vert$, of a controlled transport force, has to be 
interpreted dynamically. Since the acceleration force depends on the time 
duration in which the transport is performed, this requirement can be violated 
for a slower transport. In Fig.\ref{waveforms}) we show that in the initial 
phase the perturbations overwhelm the transporting acceleration for aspect 
ratios $\hat{W}\ge 1.5$ or larger. Results for smaller aspect ratios are not 
given in this figure because the transport force by far dominates the excess 
force and lead to a fully controlled transport.

In general, fluctuations in the frequency and transport force affect an energy 
transfer according to Eq.(\ref{firstorderheating}). This introduces violations 
to the symmetry of the transporting force and leads to an enhanced energy 
transfer as seen from Eq.(\ref{symmetry}). We have not included higher order 
terms in  our discussion, because we aim for experimental conditions to 
perform a transport in the well-controlled regime. However, they are 
inevitable for longer transport distances and other types of motion, such as 
nonadiabatic transport, or splitting of ion groups where they might lead to 
large energy transfers.

\section{Conclusions}
In conclusion, we have analyzed the dynamics of single ion transport in 
microstructured linear Paul trap arrays. We have modeled the transport by 
a forced and parametrically excited harmonic oscillator and have presented a 
theoretical framework for its description. We have derived exact analytical 
expressions for the classical as well as quantum dynamics and reviewed their 
related properties. In particular we have expressed the Heisenberg operators 
by the approach of Kim et al. \cite{kim} through the dynamical quantities of 
the related classical solution. We have given explicit analytical expressions 
for the classical energy transfer involved in these transport phenomena and 
derived expressions for the lowest order deviations from ideal transport that 
will necessarily appear for unfavourable ion trap layouts. For current trap 
technology we have evaluated durations for a fast adiabatic transport and 
found that they depend strongly on the external force employed in the 
transport. According to these results, the adiabatic single ion transports of 
reference \cite{rowe} could be sped up by more than an order of magnitude
with negligible energy transer to the motion. We 
determined appropriate transport waveforms and found that with an adiabatic 
transport over four electrode stripes of size roughly equal to the distance of 
the ion to the nearest electrode and frequencies in the range of $\sim 
9~\text{MHz}$ is feasible in about 6 oscillation cycles. Our results also 
indicate that a full control over the transport is available, where 
perturbations to a harmonic oscillator potential are negligible at all 
positions and times. By directly relating deviations from these ideal 
potentials to the aspect ratio of the trap, we have found a practical design 
rule that should be valid for trap layouts more general than the one given 
here. The ratio of a control electrode width to the distance to the ion should 
be in the range $0.5-1$ for a well-controlled regime. Our example suggests 
that a higher electrode density does not appreciably improve transport 
performances. This provides important insight into the amount of resources 
needed to realize large scale implementations of ion trap based quantum 
computers. Transport in a confining well of constant frequency might also 
enable continuous cooling processes during the transport. If eventually 
experiments allow one to maintain a well-controlled regime during the 
transport, performing quantum processing during transport is conceivable, 
possibly leading to appreciably shorter processing times.

\section{Acknowledgements}
R.R. acknowledges support by the Alexander von Humboldt Foundation during the 
course of this work.  Work also supported by DTO and NIST.
We thank A. Steane, T. Rosenband and J. Wesenberg for helpful comments 
on the manuscript.
\appendix

\section{Integral expansion}\label{integralexpansion}
We employ a mathematical theorem proven within the formalism of 
h-transforms, see for example theorem 3.2 of \cite{bleistein}:
If $g(\tau)$ has $N+1$ continuous derivatives while $g^{(N+2)}$ is piecewise 
continuous on the real axis $[a,b]$ then
		\begin{equation}
	I(\lambda)= \int_a^b e^{-i \lambda \tau} g(\tau) d\tau 
	  \stackrel{\lambda\rightarrow\infty}{\sim} 
	  \sum_{n=0}^N \frac{(-1)^n}{(-i\lambda)^{n+1}}
	   \left[g^{(n)}(b)e^{-i\lambda b}-g^{(n)}(a)e^{-i\lambda a}\right].
	  \end{equation}
If we also require 
$\lim_{\tau\rightarrow a+} g^{(n)}(\tau)=\lim_{\tau\rightarrow b-} 
g^{(n)}(\tau)=0$ for $n=0,..,N-1$ 
it holds that $I(\lambda)=o(\lambda^{-N})$.


\end{document}